\documentstyle[aps,prb]{revtex}
\input epsf
\begin{document}

\twocolumn[\hsize\textwidth\columnwidth\hsize\csname@twocolumnfalse\endcsname

\title {Charged Vortices in High Temperature Superconductors Probed by NMR 
}

\author{ Ken-ichi Kumagai$^{1,3}$,  Koji Nozaki$^{1}$ and Yuji 
Matsuda$^{2,3}$}
\address{$^{1}$ Division of Physics, Graduate School of Science, Hokkaido
University, Kita-ku, Sapporo 060-0810, Japan}
\address{$^{2}$ Institute for Solid State Physics, University of Tokyo,
Kashiwanoha, 5-1-5, Kashiwa, Chiba 277-8581, Japan}
\address{$^{3}$ CREST, Japan Science and Technology Corporation, 
Kawaguchi, Saitama
332-0012, Japan }

\date{received ~~~~~~~~~~~~~~~~~~~~~~~~~~~~~~~}
\maketitle

\begin{abstract}
We report a first experimental evidence that a vortex in the high temperature superconductors (HTSC) traps a finite electric charge from the high resolution measurements of the nuclear quadrupole frequencies. In slightly overdoped YBa$_2$Cu$_3$O$_7$ the vortex is negatively charged by trapping electrons, while in underdoped YBa$_2$Cu$_4$O$_8$ it is positively charged by expelling electrons. The sign of the trapped charge is opposite to the sign predicted by the conventional BCS theory. Moreover, in both materials, the deviation of the magnitude of the charge from the theory is also significant. These unexpected features can be attributed to the novel electronic structure of the vortex in HTSC.

\end{abstract}

]

\narrowtext
\section {Introduction}
	One of the most important physical properties of the vortex created
in the type-II superconductors is that a vortex line can support a magnetic
flux with a flux quantum
$\Phi_0=hc/2e(=2.07\times10^{-7}$~Oe$\cdot$cm$^{2}$) \cite{degennes}.   
This
fact has been
confirmed experimentally more than 4 decades ago.  On the other hand, it is
only very recent that another prominent feature, namely the possibility 
that
a vortex of the superconductor can accumulate a finite electric charge as
well, has come to be realized \cite{khomskii,blatter,mishonov,hayashi}.
The vortex charge appears as a result of the chemical potential difference
between the vortex core and the region away from the vortex core. It 
should be 
emphasized that the sign and magnitude of this charge is closely related 
to the 
microscopic electronic structure of the vortex, which in turn reflects all 
the
fundamental natures of the superfluid electrons and the low energy
excitation out of the condensate.  Moreover,  it has also been pointed out
recently that the vortex charge strongly affects the dynamical properties 
of
the vortex \cite{dorsey,otterlo,feigelman,kato,eschrig}.  For instance, the
origin of the vortex Hall anomaly  has been attributed to the vortex 
charge.
Though the clarification of the issue of the vortex charge serves as an
important test of the predictions for the vortex electronic structure and
the dynamics, it has never been examined experimentally so far.   This is
mainly because in conventional superconductors the magnitude of the
accumulated charge within the core is very small and is extremely difficult
to observe.

	In this paper we report a first straightforward attempt to
identify the vortex charge in the high temperature superconductors (HTSC) 
by
high resolution measurements of the nuclear quadrupole frequency $\nu_Q$,
which is very sensitive to the local charge density.  We show that a vortex
in HTSC indeed traps a small but finite electronic charge as well.   In
slightly overdoped YBa$_2$Cu$_3$O$_7$ the vortex is negatively charged,
while in underdoped YBa$_2$Cu$_4$O$_8$ it is positively charged.  The sign
of the trapped charge is opposite to the sign predicted by the 
conventional BCS theory.    
Moreover, in both materials, the accumulated charge is much larger than 
expected in the ordinary superconductors.  We discuss several possible 
origins for these
discrepancies.

\section {vortex charge}
	   We first briefly introduce the vortex charge in type-II
superconductors.  In the core of the conventional superconductors, the
distance between each discrete energy levels of the quasiparticles in the
Andr\'{e}ev bound states are merely of the order of a few mK.  Thus
it is sufficient to view the energy levels as forming a continuous 
spectrum,
just like in a normal metallic state \cite{degennes,gygi}. Generally the 
chemical
potential $\mu$ in the superconducting state differs from that in the 
normal
state, if an electron-hole asymmetry is present. Assuming therefore that 
the
vortex core is a normal metallic region surrounded by the superconducting
materials, this difference in $\mu$ is expected to arise and should lead to
the redistribution of the electrons. In order to maintain the same
electrochemical potential on both side, the charge transfer occurs between
the core and the outside.

	   In the framework of the BCS theory taking into account the
metallic screening effect, the charge accumulated within the vortex core
$Q_{\xi}$ per layer normal to the magnetic field is given as,
 \begin{equation}
  Q_{\xi} \approx
\frac{2ek_Fs}{\pi^3}\left(\frac{\lambda_{TF}}{\xi}\right)^2
  \left(\frac{d\ln T_c}{d\ln\mu}\right),
  \end{equation}
where $\lambda_{TF}$ is the Thomas-Fermi screening length, $s$ is the
interlayer distance, $\mu$ is the chemical potential and $e(>0)$ is the
electron charge \cite{blatter}. The sign of the core charge is determined
by $d\ln T_c/d\ln\mu$, which represents the electron-hole asymmetry.
Outside the core, the charges with opposite sign screen the core charge,
similar to a charged particle in a metal. Far outside the core, this
screening charges decay gradually with a power law dependence as $r^{-4}$
(see Fig.1 (a)) \cite{blatter,kato}.  In strong fields ($H_{c1} \ll H \ll
H_{c2}$),  
\begin{figure}
 \centerline{\epsfxsize 8cm \epsfbox{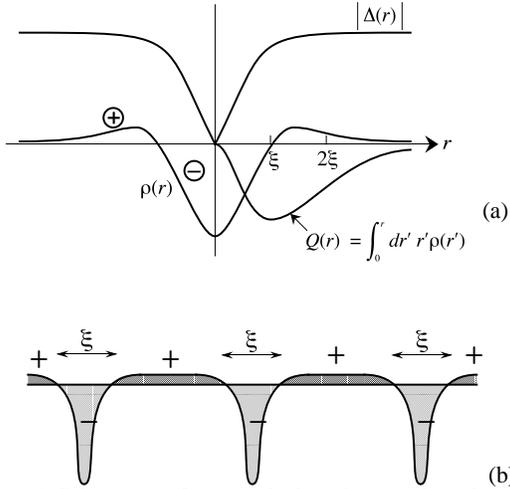}}
\caption{{\bf (a)}:Schematic figure of the charge distribution around a 
single vortex core when the electrons are trapped within the core 
(negatively charged core).
$\Delta(r)$ is the superconducting energy gap.   $\rho(r)$ is the charge
density.  The charge accumulated inside the core are screened by the 
charges
with opposite sign.  $\rho(r)$ decays gradually as $r^{-4}$ well outside 
the
core region.  $Q(r)$ is the total charge within the distance $r$.  $Q(r)$
goes to zero as $r \rightarrow \infty$ due to the requirement of the 
overall
charge neutrality. {\bf (b)}: The charge density modulation in the strong 
magnetic
field ($H_{c1} \ll H \ll H_{c2}$) where each vortices overlaps.   }
\end{figure}
\noindent each vortex overlaps with its neighborhood.  Then the charge
density outside the core is nearly constant, and the periodic modulation of
the charge density appears for the periodic vortex lattice (Fig.1(b)).  In
ordinary superconductors, $|Q_{\xi}|$ is estimated to be
$\sim10^{-5}-10^{-6}e$, using $k_F\sim1$\AA$^{-1}$,   $\lambda_{TF}\sim
k_F^{-1}\sim1$\AA, $\xi \sim 100 $\AA~and  $|d \ln T_c/d\ln \mu  |\approx
\ln (\hbar\omega_D/k_BT_c)  \sim 1-10$, where $\omega_D$ is the Debye
frequency.  Thus $|Q_{\xi} |$ is negligibly small and is very difficult
to observe.

	However, the situation in the case of HTSC seems to be promising
because $\xi$ is extremely short compared to that of the conventional
superconductors.  Moreover the strong electron correlation effects and
$d$-wave pairing symmetry of HTSC are expected to change the electronic
structure of the vortex dramatically
\cite{schopohl,morita,arovas,himeda,franz,andersen}.  In fact, recent STM
measurements revealed that the vortex of HTSC is very different from those
of conventional superconductors \cite{fischer,renner,pan}.  These unusual
features of HTSC are expected to enhance the charging.  We will discuss
these issue later.

	The vortex charge also plays an important role for the vortex
dynamics.  When a vortex moves in the superfluid electrons, the core plays
a key role in dissipation processes \cite{eschrig}.  One of the most
striking phenomena is the vortex Hall anomaly, namely sign reversal of the
flux flow Hall effect below $T_c$, which is observed in most HTSC
\cite{nagaoka}. This Hall sign reversal indicates that the vortices move
upstream against the superfluid flow.  Such an unusual motion has never 
been
observed in any other fluid including superfluid helium and cannot be
explained in the 
\begin{figure}
\centerline{\epsfxsize 7cm \epsfbox{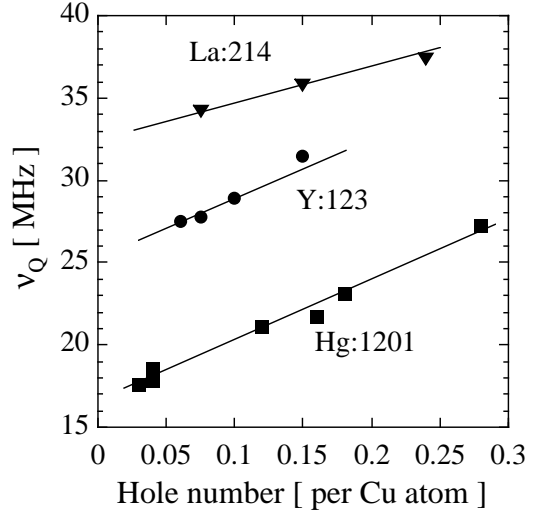}}
\caption {Doping dependences of $\nu_Q$ on the Cu(2) site for 
YBa$_2$Cu$_3$O$_{7-\delta}$, \cite{yasuoka} La$_{2-x}$Sr$_x$CuO$_4$,  
\cite{zheng} and HaBa$_2$CuO$_{4+\delta}$. \cite{gippius} $\nu_Q$ is 
proportional to the hole number.}
\end{figure}
\noindent framework of the classical hydrodynamic theory
\cite{bardeen}.  Recently, this phenomena has been discussed in terms of
the vortex charge which produces an additional force acting on the
vortices \cite{dorsey,otterlo,feigelman,kato}.

   To elucidate the vortex charge, a direct observation of the local 
carrier
density is strongly required.  It has been pointed out by many authors that
$\nu_Q$ in HTSC is very sensitive to the local hole density
\cite{schwarz,yasuoka,zheng,gippius}. In a solid, an electron distribution
with spherical asymmetry such as unclosed 3$d$ shell and noncubic
surrounding ions induce a local electric field gradient (EFG) in the
vicinity of the nuclei. This EFG lifts the degeneracy of the nuclear spin
levels, interacting with the nuclear quadrupole moment, $Q_N$.  The
relevant information is obtained from the nuclear quadrupole resonance
(NQR) in zero magnetic field and the nuclear magnetic resonance (NMR) in a
finite magnetic field.  For the $^{63}$Cu nuclei with spin $I$=3/2, the NQR
resonance frequency $\nu_Q^{NQR}$ is expressed as
\begin{equation}
     \nu_Q^{NQR}=\frac{e^{2}Q_{N}q_{zz}}{2h} \sqrt{1+\frac{\eta^2}{3}}
              =\nu_Q\sqrt{1+\frac{\eta^2}{3}}
 \end{equation}
where $eq_{zz}$ is the largest principle ($z-$axis) component of EFG at the
nuclear site, $Q_N$ (=$-$0.211 barn for $^{63}$Cu) is the quadrupole  
moment
of copper nuclei.\cite{carter} The asymmetry parameter $\eta$ of the EFG
defined as
$\eta=|(q_{xx}-q_{yy})/q_{zz}|$  is close to zero for the Cu site in the
two dimensional CuO$_2$ planes (Cu(2) site).

     In a strong magnetic field when the Zeeman energy is much larger than 
the
quadrupole energy, each Zeeman level is shifted by the quadrupole
interactions and thus, two satellite peaks ($\pm 3/2\leftrightarrow \pm
1/2)$ appear on both sides of the central ($\pm 1/2\leftrightarrow \mp 
1/2$)
resonance peak. The frequency difference between the upper and the lower
satellites exactly coincides with 2$\nu_Q$, when a magnetic field is
applied parallel to the largest  principle axis of EFG, namely,
$H\parallel c$-axis in the present experimental 
\begin{figure}
\centerline{\epsfxsize 9cm \epsfbox{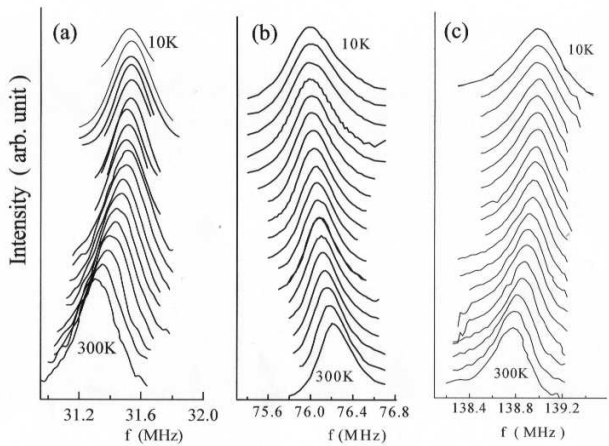}}
\caption{The NQR (a), the lower (b) and the upper (c) NMR satellite 
spectra at 9.4T for $^{63}$Cu(2) site of YBa$_2$Cu$_3$O$_7$ for various 
temperatures. (220K, 200K, 180K and 160k to 10K by a 10K step.)  }
\end{figure}
\noindent condition described below. 
Even for the second order or more higher order quadrupole effect in the 
presence of the asymmetry $\eta$ of EFG, the each satellite line shifted 
as much and the corrections for 2$\nu_Q$ are vanishing in the case of 
$H\parallel c$-axis. \cite{carter}.
 
   Generally the EFG originates from two different sources, namely from the
on-site distributions $q_{on-site}$ of the electrons and from the
surrounding ions $q_{ion}$, $q=q_{on-site}+q_{ion}$.  Recent analysis of 
$q$
on the Cu(2) site suggests that $q_{on-site}$ is mainly composed of the Cu
4$p$ and 3$d$ shell terms \cite{schwarz}. In HTSC the holes in the Cu 
3$d_{x^2-y^2}$ 
orbital play an important role for the onset of superconductivity. Figure 
2 shows the doping dependence of $\nu_Q$ of the Cu(2) site for
YBa$_2$Cu$_3$O$_{7-\delta}$ \cite{yasuoka}, 
La$_{2-x}$Sr$_x$CuO$_4$ \cite{zheng}, and 
HaBa$_2$CuO$_{4+\delta}$.\cite{gippius} 
In all materials, $\nu_Q$ increases linearly with the number of holes in 
the planes and can be written as,
\begin{equation}
\nu_Q=An_{hole}+C,
\end{equation}
where $n_{hole}$ is the number of holes per Cu(2) atom, and $A$ and $C$ are
constants \cite{yasuoka,zheng,gippius}.  Although  $C$ is strongly material
dependent, reflecting the difference in $\nu_{ion}$,  $A$ 
$\approx20-30$~MHz
per hole for the Cu(2) atom is essentially material independent.  Thus, the
precise measurement of $\nu_{Q}$ makes possible an accurate determination 
of
the change of the local hole number at the Cu site.

\section {Experimental}
 The principle of our experiment is the following.  In the measurement,
only the resonance of the $^{63}$Cu(2) nuclei {\it outside the vortex core}
is detected.  This is because the applied field is much less than $H_{c2}$
and hence the core region occupies a smaller area in the sample.  If the
vortex core traps (expels) a finite amount of electrons, the electron
\begin{figure}
\centerline{\epsfxsize 9cm \epsfbox{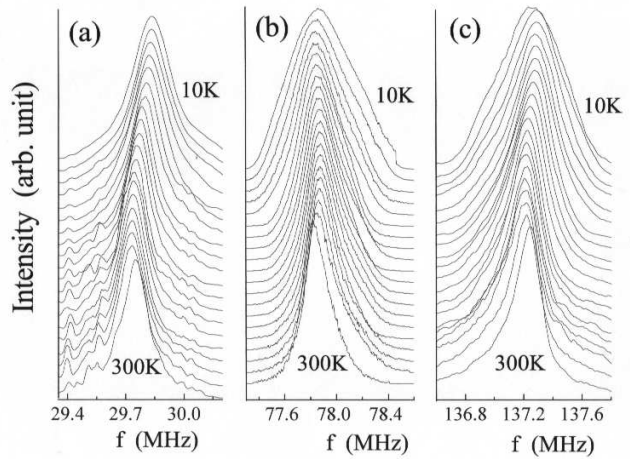}}
\caption{ The NQR in zero field (a) and the lower (b) and the upper (c)
NMR-satellite spectra at 9.4T for $^{63}$Cu(2) site of YBa$_2$Cu$_4$O$_8$ 
for various
temperatures. (300K, 220K, 200K, 180K and 160k to 10K by a 10K step.)  }
\end{figure}
\noindent density outside the core should decrease (increase) from that in zero field
where the electron distribution is uniform, as shown in Fig.1(b).  We are
able to detect the change of carrier density through the change of the 
value
of  $\nu_Q$. 

In the present measurements, we used slightly overdoped
YBa$_2$Cu$_3$O$_7$ and underdoped YBa$_2$Cu$_4$O$_8$ in which the NQR and
NMR spectra are very sharp compared to those of other HTSC.  The NMR 
spectra
are obtained for fine powdered sample (the grains are less than 33 $\mu m$)
with uni-axial alignment. The each grain aligns to an easy axis (the
$c$-axis) in a high external field at room temperature, and then
fixed with epoxy (Stycast 1266).

 The Cu-NQR and NMR spectra were obtained by a conventional pulse 
spectrometer. The NMR experiments were performed in the field cooling 
condition under a constant field of 9.4T by using a highly-homogeneous 
superconducting magnet which was stabilized to less than 1ppm during the 
experiment. The present measurement in the vortex state is made in the 
so-called Bragg-glass phase in which the quasi-long-range order of the 
vortex lattice is preserved. We see slightly-asymmetric shape of the 
NMR-satellite spectra due to the somehow miss-alignment of some grains. 
The line broadenings of the NQR spectra in zero external field are not 
observed even below $T_c$ in both YBa$_2$Cu$_3$O$_7$ and 
YBa$_2$Cu$_4$O$_8$.  The width of the NMR spectra becomes broader with 
decreasing $T$ below $T_c$ partly due to the magnetic field inhomogeneity 
caused by the introduction of the vortices and partly due to the spatial 
distribution of carrier density, which we will discuss later. 
\begin{figure}
\centerline{\epsfxsize 6cm \epsfbox{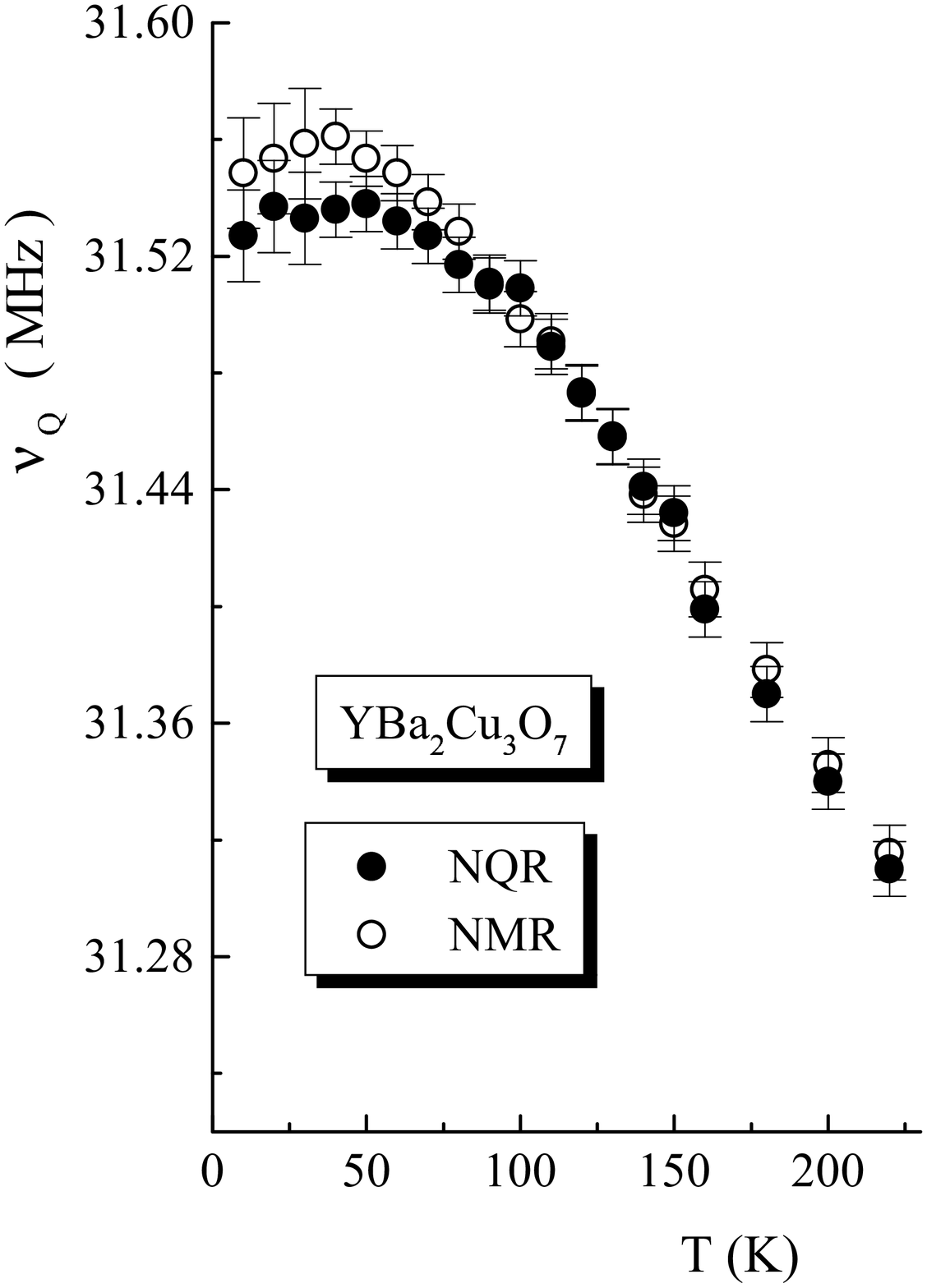}}
\caption{ The temperature dependence of $\nu_Q$ obtained from NQR and NMR 
for $^{63}$Cu(2) site of YBa$_2$Cu$_3$O$_7$.  }
\end{figure}
\section {results and discussion}

 Figure 3 shows the NQR and lower and upper satellite spectra of $^{63}$Cu(2) NMR 
for optimum doped YBa$_2$Cu$_3$O$_7$.  Figure 4 shows also the NQR and NMR 
spectra of $^{63}$Cu(2) site of YBa$_2$Cu$_4$O$_8$.  All data were taken 
by utilizing a phase coherent spin echo spectrometer.  The spectra are 
obtained with a superposed method of the Fourier Transform spectra of the 
spin echo measured at a certain
frequency interval. The temperature dependences of $\nu_Q$ obtained fron 
NMR and NQR are shown for YBa$_2$Cu$_3$O$_7$ and YBa$_2$Cu$_4$O$_8$ in 
Figs. 5 and 6, respectively. The temperature dependences of $\nu_Q^{NQR}$ 
for both YBa$_2$Cu$_3$O$_7$ and YBa$_2$Cu$_4$O$_8$ are quite similar to 
those of $\nu_{Q}^{NQR}$ reported previously \cite{brinkmann}. It should 
be noted that the procedure for obtaining $\nu_Q$ from the frequency 
difference between the upper and the lower satellites is essentially free 
from the influence of the change of magnetic shift (or Knight shift). 
Moreover, the magnetic effect of the asymmetric broadening (so-called 
Redfield pattern) due to the vortex lattice in the superconducting state 
is exactly cancelled out in the process determining $\nu_Q$ from NMR. 
Thus, we obtained the $\nu_Q$ values simply from the difference of the 
peak frequencies of the two satellite lines.

We plot in Fig. 7 the difference between $\nu_Q$ in zero field and in the
vortex state, $\Delta\nu_Q=\nu_Q(0)-\nu_Q(H)$, for YBa$_2$Cu$_3$O$_7$ and
YBa$_2$Cu$_4$O$_8$.  The $\nu_Q$ in zero field for NQR is obtained after 
the
correction by the factor of $\sqrt{1+\eta^2/3}$ in Eq.(1), although this
factor is at most 0.03\% of $\nu_Q$ for $\eta\approx$0.04 of the present
materials.  In both materials $\Delta\nu_Q$ is essentially zero above 
$T_c$,
indicating no modulation of the carrier density.  Meanwhile a 
\begin{figure}
\centerline{\epsfxsize 6cm \epsfbox{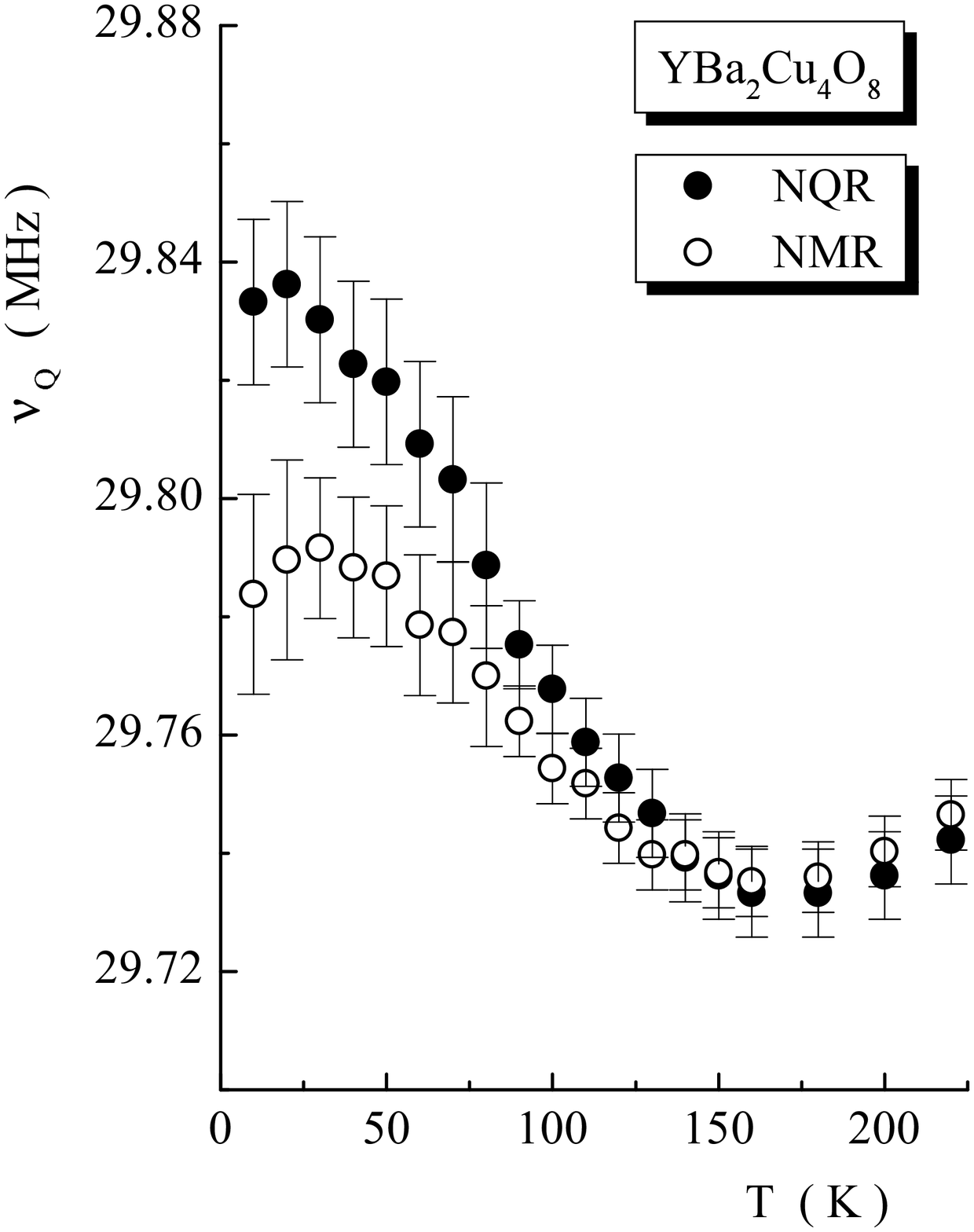}}
\caption{ The temperature dependence of $\nu_Q$ obtained from NQR and NMR 
for $^{63}$Cu(2) site of YBa$_2$Cu$_4$O$_8$.  }
\end{figure}
\noindent nonvanishing
$\Delta\nu_Q$ is clearly observed below $T_c$ in both materials.  While
$\Delta\nu_Q\sim-25$kHz is of negative sign and in YBa$_2$Cu$_3$O$_7$,
$\Delta\nu_Q\sim50$kHz is of positive sign in YBa$_2$Cu$_4$O$_8$ at $T$=0.

   We discuss here several possible origins for the nonzero $\Delta\nu_Q$.  We first point out that the magnetostriction cannot be the origin of the nonzero $\Delta\nu_Q$.     In fact, both the magnetostriction and ultrasonic absorption measurements showed that the local lattice distortion $\Delta \ell$ caused by the magnetostriction is negligibly small under the field cooling condition; $\Delta \ell/\ell < 10^{-8}$ below 10~T where $\ell$ is the lattice constant.   We next remark on an in-plane charge modulation caused by a charge stripe formation \cite{tranquada} or the charge density wave  (CDW) transition of the chain site.\cite{kramer,grevin}  Although the static charge ordering associated with a stripe formation has been discussed in some of the high-$T_c$ cuprates, such orderings have never been reported in YBa$_2$Cu$_3$O$_7$ nor in YBa$_2$Cu$_4$O$_8$.  Moreover, it has been reported that in La:214 compounds the static charge stripe order gives rise to the wipe-out of NQR signals\cite{hunt}, implying that the static stripe order, if present, induces a large effects on quadrupole interactions.  However, we have not observed the broadening of the NQR spectra (in zero field) and nor any sign of the wipe-out of NQR and NMR signals in the present measurements for YBa$_2$Cu$_3$O$_7$ or YBa$_2$Cu$_4$O$_8$.    Finally we mention a recent report of the broadening of the NQR line widths in the plane at low temperatures in YBa$_2$Cu$_3$O$_{7-\delta}$, which was explained in terms of the CDW formation in the chain site.\cite{grevin}    In the present study, in which we obtained sharper NQR line widths compared to the previous reports \cite{kramer,grevin}, such line-broadenings due to the electric quadrupole interaction is not observed.  Thus there is no evidence of the charge modulation due to CDW or stripe formation  
\begin{figure}
\centerline{\epsfxsize 7cm \epsfbox{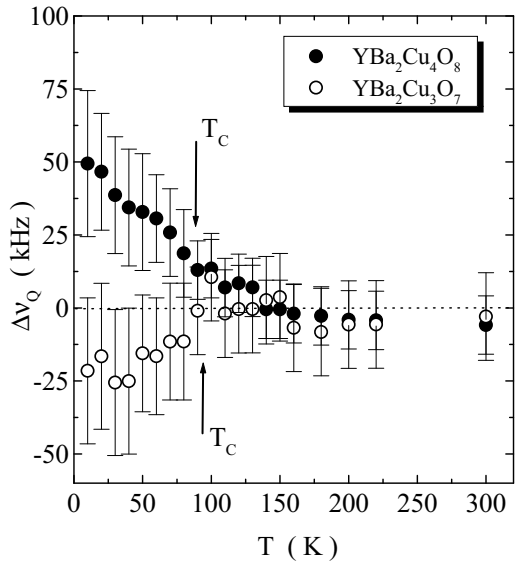}}
\caption{$T$-dependence of $\Delta\nu_Q=\nu_Q(0)-\nu_Q(H=9.4~T)$ for 
YBa$_2$Cu$_3$O$_7$ and YBa$_2$Cu$_4$O$_8$.  In both materials nonzero
$\Delta\nu_Q$ is clearly observed below $T_c$, showing that the electron
density outside the core differs from that in zero field. }
\end{figure}
\noindent  in zero field in our crystals.  It is also quite unlikely that the magnetic field induces the CDW or stripe formations.  Having ruled out these various possibilities,  we conclude that the nonzero values for $\Delta\nu_Q$ naturally lead to the fact that {\it the electron density outside the vortex core is different from that in zero field.}

  We now discuss the issue of the sign and magnitude of the accumulated
charge.  The negative $\Delta\nu_Q$  in YBa$_2$Cu$_3$O$_7$ indicates the
increment of the hole density outside the core.  This excess density of
holes is nothing but the holes expelled from the core. Therefore the
accumulated charge in the core of YBa$_2$Cu$_3$O$_7$ is negative.  By the
same reasoning, the positive $\Delta\nu_Q$ in YBa$_2$Cu$_4$O$_8$ indicates 
a positive accumulated charge. Meanwhile, since the chemical potential
decreases monotonically with doping holes,  Eq.(1),  sgn$Q_{\xi}=$sgn$(d\ln
T_c/d\ln\mu)$,  predicts that $Q_{\xi}>0$ in the underdoped regime while
$Q_{\xi}<0$ in the overdoped regime. This is strikingly in contrast to the
sign determined by the present experiment. The deviation of the magnitude
of the charge from theory is also noteworthy. The magnitude of charges per
pancake vortex, which are roughly estimated using $Q_{\xi}\approx\Delta
\nu_QH_{c2}/AH$ assuming $H_{c2}\sim200$~T are $Q_{\xi}\sim$ -0.005$e$ to 
-0.02$e$
for YBa$_2$Cu$_3$O$_7$ and  $Q_{\xi}\sim$0.01$e$ to 0.05$e$ for 
YBa$_2$Cu$_4$O$_8$.
However, according to Eq.(1),  $Q_{\xi}$ is estimated to be
$\sim10^{-4}-10^{-5}e$ where we assumed $\xi\sim 30\AA$.   Therefore
$|Q_{\xi}|$ determined by the present experiments are still one or two 
order
of magnitude larger than expected by Eq.(1). Thus {\it the BCS theory not
only predicts the wrong sign of the charge but also underestimates
$|Q_{\xi}|$ seriously.}

     There are several intriguing possible origins for these discrepancies.
For example, because of the extremely short $\xi$, the vortex core may be 
in the quantum limit $k_F\xi\sim$1, where $k_F$ is the Fermi wave number. 
In this limit, the description of the quasiparticles in terms of 
semiclassical wavepackets breaks down in contrast to conventional 
superconductors \cite{morita}.  Furthermore, as suggested by recent 
theories of the vortex core based on {\it e.g.} the $t$-$J$ or SO(5) 
models,  the antiferromagnetic (AF) state may be energetically preferable 
to
the  metallic state in the vortex core of HTSC 
\cite{arovas,himeda,andersen}. 
 If this is indeed so, the AF correlation is expected to enhance the 
charging effect because it causes a large shift of $\mu$ by changing the 
density of states of the electrons inside the core dramatically.  We note 
here that the present results exclude the possibility of the SO(5) 
insulating AF core \cite{arovas,andersen} in which holes should be 
expelled from the core and the accumulated charges are always negative; 
the present result yields the opposite sign for YBa$_2$Cu$_4$O$_8$ in the
underdoped regime where the AF correlation is important. Therefore a 
detailed microscopic calculation is needed to evaluate the accumulated 
charge quantitatively including the sign.

\section {conclusion}

From the precise $^{63}$Cu-NMR and NQR measurements, we have shown that a 
vortex in type-II superconductors can trap a finite electric charge as 
well as magnetic flux. In the slightly overdoped YBa$_2$Cu$_3$O$_7$ the 
vortex is negatively charged, while in underdoped YBa$_2$Cu$_4$O$_8$ it is 
positively charged.
In both high $T_c$ materials, the accumulated charges are much larger than 
expected in the ordinary superconductors. The sign and value of the 
charges indicate the novel electronic structure of the vortex in HTSC. 

\section {acknowledgments}
We thank Y.~Kato, K.~Maki, T.~Mishonov, M.~Ogata,  D.~Rainer, E.B.~Sonin,
M.~Takigawa, A.~Tanaka,  and H.~Yasuoka for helpful discussions.  We also
thank H.~Ikuta and T.~Hanaguri for comments on the magnetostriction. 
We also thank S. Shamoto and T. Isobe for their technical assistance on 
the crystal growth and K.~Kakuyanagi for his assistance on NMR 
measurement. This work was supported by the REIMEI Research Resources of 
the Japan Atomic
Energy Research Institute, Iketani Science and Technology Foundation and a
Grant-in-Aid for Scientific Research on Priority Area "Vortex Electronics"
from the Ministry of Education, Science, Sports and Culture of Japan.

\end{document}